\documentclass[runningheads]{llncs}
\usepackage{graphicx}

\usepackage{extdash}

\usepackage[utf8]{inputenc}
\usepackage[T1]{fontenc}

\usepackage{cite}
\usepackage{xurl}
\usepackage{hyperref}

\usepackage{color}

\usepackage{caption}
\makeatletter
\newcommand\notsotiny{\@setfontsize\notsotiny\@vipt\@viipt}%
\makeatother

\usepackage{circledsteps}
\pgfkeys{/csteps/inner color=white}
\pgfkeys{/csteps/outer color=black}
\pgfkeys{/csteps/fill color=black}

\usepackage{multirow}
\usepackage[utf8]{inputenc}

\usepackage{booktabs}

\usepackage{xurl}

\usepackage{eso-pic}

\AddToShipoutPictureBG*{
  \AtPageCenter{
    \raisebox{11.5cm}{
      \makebox[0pt]{
        \fbox{\parbox{0.95\textwidth}{\small This paper has been accepted for publication at DIMVA 2025. If you wish to cite this work, please use the following reference: Hiroki Nakano, Takashi Koide, and Daiki Chiba, ``ScamFerret: Detecting Scam Websites Autonomously with Large Language Models,'' Proceedings of the 22nd Conference on Detection of Intrusions and Malware and Vulnerability Assessment (DIMVA 2025).}}
      }
    }
  }
}
\begin{document}
\title{ScamFerret: Detecting Scam Websites Autonomously with Large Language Models}
\titlerunning{ScamFerret}
\authorrunning{H. Nakano et al.}
\author{Hiroki Nakano \and 
Takashi Koide \and
Daiki Chiba}
\institute{NTT Security Holdings Corporation \& NTT Corporation, Tokyo, Japan \email{hi.nakano.sec@gmail.com}}
\maketitle              %
\begin{abstract}
With the rise of sophisticated scam websites that exploit human psychological vulnerabilities, distinguishing between legitimate and scam websites has become increasingly challenging.
This paper presents ScamFerret, an innovative agent system employing a large language model (LLM) to autonomously collect and analyze data from a given URL to determine whether it is a scam.
Unlike traditional machine learning models that require large datasets and feature engineering, ScamFerret leverages LLMs' natural language understanding to accurately identify scam websites of various types and languages without requiring additional training or fine-tuning.
Our evaluation demonstrated that ScamFerret achieves 0.972 accuracy in classifying four scam types in English and 0.993 accuracy in classifying online shopping websites across three different languages, particularly when using GPT-4.
Furthermore, we confirmed that ScamFerret collects and analyzes external information such as web content, DNS records, and user reviews as necessary, providing a basis for identifying scam websites from multiple perspectives.
These results suggest that LLMs have significant potential in enhancing cybersecurity measures against sophisticated scam websites.
\keywords{Large Language Model \and Human Psychological Vulnerability \and Scam Website \and Agent}
\end{abstract}
\section{Introduction}
Scam websites have become an increasingly prevalent threat, causing significant financial losses and personal information compromise.
In 2023, reported losses in the United States reached \$12.5 billion, a 22\% increase from the previous year~\cite{fbi_report}.
While traditional phishing websites often mimic legitimate websites and can be detected through specific visual cues~\cite{DBLP:conf/uss/LinLDNCLSZD21,DBLP:conf/IEEEares/DooremaalBAZ21}, modern scam websites have evolved to become highly sophisticated, making them challenging to identify even for security experts.
These sophisticated scam websites exploit human psychological vulnerabilities, perpetuating deception and escalating financial losses, evolving into a significant societal issue that demands urgent attention.

Previous research has focused on developing machine learning models to detect scam websites using HTML content and domain name information~\cite{DBLP:conf/ndss/LiYN23,DBLP:conf/www/SrinivasanKMANA18}.
However, these approaches face several limitations:
\begin{itemize}
    \item They require large labeled datasets for each scam type and language, which are time-consuming and costly to create.
    \item They demand complex feature engineering specific to each scam variant, limiting generalizability.
    \item They lack transparency in the detection process, making it difficult for users to understand the basis for decisions intuitively.
\end{itemize}

To address these challenges, we present ScamFerret, a novel agent-based system for analyzing diverse scam websites across multiple languages without requiring additional training on scam-specific data.
ScamFerret leverages a large language model (LLM) that has already been trained on a broad corpus of Internet text, which likely includes some information about scams and fraudulent activities.
This pre-existing knowledge allows the system to operate effectively without scam-specific fine-tuning.

ScamFerret uses the LLM to autonomously select appropriate information-gathering tools, collect useful information for website analysis, and perform contextual analysis to identify scam websites.
By utilizing the natural language understanding capabilities and broad knowledge base of LLMs, ScamFerret can recognize subtle suspicious elements and provide explanations for its classifications, drawing on its general understanding of language, web content, and potential fraudulent patterns.

We evaluate ScamFerret on new datasets comprising four types of scam websites (fake online shopping, technical support scams, cryptocurrency scams, and investment scams) in English, as well as three languages (English, German, and Japanese) for fake online shopping websites.
Our results demonstrate that ScamFerret achieves a mean classification accuracy of 0.972 across four scam types in English and 0.993 across three languages for online shopping websites when using GPT-4, outperforming both conventional machine learning-based detectors~\cite{DBLP:conf/sp/BitaabCOLWAWBSD23,DBLP:conf/acsac/KotziasRPSB23} and simpler LLM-based approaches.

This paper makes the following contributions:
\begin{itemize}
\item We introduce ScamFerret, a system that autonomously collects and analyzes data to detect scam websites without requiring large, labeled datasets for each scam type, leveraging LLMs to recognize sophisticated scam websites.
\item We demonstrate ScamFerret's effectiveness across multiple scam types and languages, achieving state-of-the-art accuracy of 0.972 for four scam types in English and 0.993 across three languages for online shopping websites using the GPT-4 model.
\item We provide an analysis of ScamFerret's detection process, offering insights into how LLMs can be leveraged for explainable web security tasks.
\item We share the code for the proposed system, the dataset used for evaluation, and the evaluation results at \url{https://github.com/ScamFerret/artifact}.
\end{itemize}

\section{Scope and Goal}
\subsection{Scope of Scam Websites}
\label{subsec:scope}
This study focuses on four types of scam websites across three languages: English, German, and Japanese.
We target fake online shopping, technical support scams, cryptocurrency scams, and investment scams, which have high victim rates and have been the subject of previous detection efforts~\cite{DBLP:conf/eurosp/LiuPVP23,DBLP:journals/corr/abs-2401-09824,DBLP:conf/www/SrinivasanKMANA18,DBLP:conf/acsac/KotziasRPSB23}.
These scams pose significant financial risks, with cryptocurrency and investment scams resulting in billions of dollars in losses annually, while online shopping scams exploit the rapidly expanding e-commerce market, leading to widespread consumer victimization~\cite{scam_reason_1,scam_reason_2}.
Our language selection addresses the global nature of online scams~\cite{fraud_distribtuion}: English, the language with the highest number of reported fraud victims globally; German, the native language of Germany, which is frequently targeted for online fraud; and Japanese, the native language of Japan, which is the most affected language in information theft.
This diverse set allows us to assess ScamFerret's effectiveness across different contexts.
The characteristics of each scam type are as follows:

\noindent\textbf{Fake Online Shopping.}
These websites mimic legitimate online shopping platforms, often advertising rare or discounted products.
They use search engine optimization and create urgency to induce purchases, resulting in financial losses for victims.

\noindent\textbf{Technical Support Scams.}
These websites falsely alert users to technical issues, prompting contact with attackers.
They often use pop-ups or fake security warnings to direct users to fraudulent support pages, where attackers may request remote access or payment for non-existent services.

\noindent\textbf{Cryptocurrency Scams.}
These websites typically employ phishing tactics through fraudulent trading platforms and wallet services.
Attackers use fake celebrity and company accounts on social media to lure potential victims.
Once users enter their credentials, attackers can steal their cryptocurrency funds.

\noindent\textbf{Investment Scams.}
These websites promise high profits or risk-free investments.
They use professional-looking designs and create urgency with limited-time offers.
Once users invest, attackers refuse refunds and eventually cease communication.

\subsection{Research Goal}
Our primary goal is to develop a system that can analyze diverse scam websites from input URLs and provide clear justification for its classifications.
As scam websites become increasingly sophisticated, conventional detection systems based on predefined blocklists and feature learning face limitations~\cite{DBLP:conf/ndss/LiYN23,DBLP:conf/www/SrinivasanKMANA18,DBLP:conf/sp/BitaabCOLWAWBSD23,DBLP:conf/acsac/KotziasRPSB23}.
We aim to address three key challenges:

\noindent\textbf{Elimination of Labeled Dataset Requirements.}
Rapidly evolving scam websites make preparing optimal labeled datasets time-consuming, requiring constant updates to keep pace with new scam tactics.
We aim to introduce a system that can analyze various scam types and languages without pre-prepared labeled datasets, adapting in real-time to emerging threats.

\noindent\textbf{Multi-type and Multi-lingual Detection with a Single Model.}
Conventional systems often use multiple models for specific scam types and languages, requiring frequent updates.
We aim to develop a single, versatile model detecting various scam websites across types and languages, eliminating specialized models and updates.

\noindent\textbf{Clear Verbalization of Detection Rationale.}
Conventional systems often lack transparency, classifying based on numerical values without clear justification.
Our approach aims to verbalize the suspicious aspects of target URLs, enhancing reliability and understanding of new scam patterns.

\section{Proposed System: ScamFerret}

\begin{figure}[!t]
    \centering
        \includegraphics[scale=0.47]{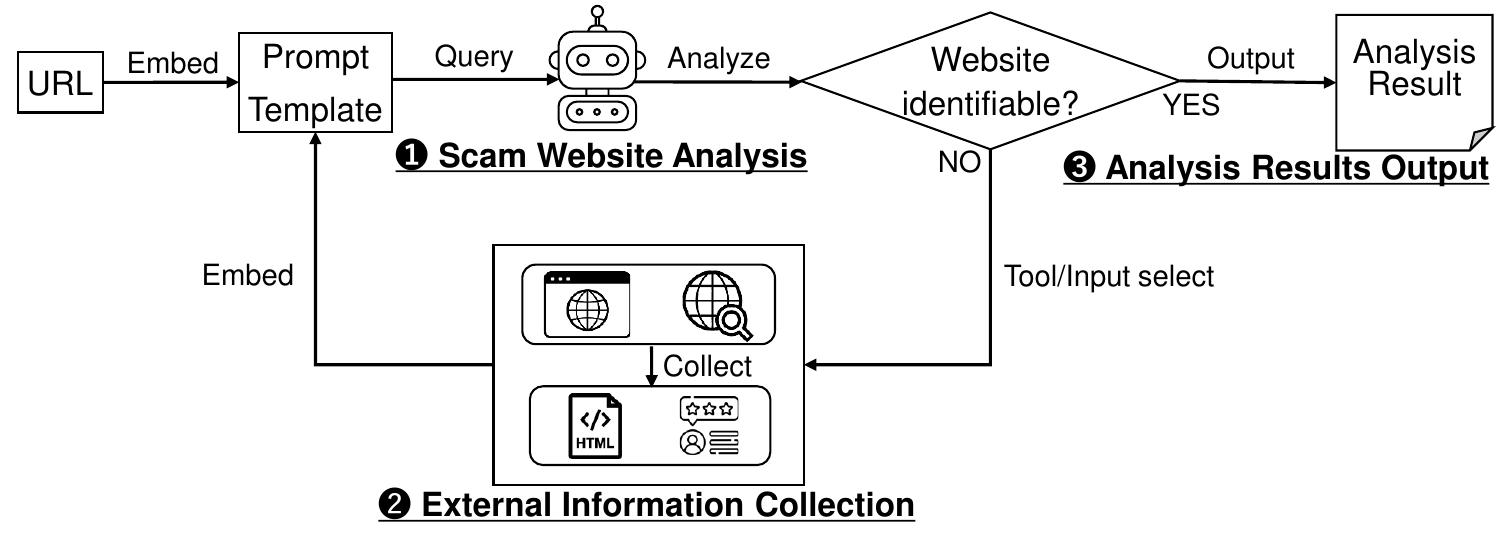}
        \caption{Overview of ScamFerret}
        \label{fig:proposed_system}
\end{figure}

\begin{table}[!t]
\tabcolsep=0.5mm
\centering
\notsotiny
\caption{Prompt Template for Proposed System}
{\renewcommand\arraystretch{0.85}
\begin{tabular}{ll} \toprule
Type & Prompt \\ \midrule
Task Setting & I want you to act as a professional scam website detection expert. \\
 & You are tasked with analyzing the content of URL given to you \\
 & to determine if the URL is a scam website or not. \\ \midrule
Characteristic Examples & Scam websites have the following features. \\
 & 1. Unusually low prices and claims of free. \\
 & 2. Claims to obtain an unusually large amount of money. \\
 & 3. Websites contain texts targeting human psychological weaknesses.\\
 & 4. Information on non-existent companies. \\
 & 5. Handling different products from common e-commerce websites. \\
 & 6. Inquiry phone number and email are unsuitable for business use. \\
 & 7. Privacy of customer information notation is ambiguous. \\
 & 8. Payment methods are not common and are unusual. \\
 & 9. The information listed has not been updated. \\ \midrule
Tool Definitions & You can access the following tools to help you answer the question: \\
 & Tool Name 1: Tool Description \textbf{(Contents of Table 2)}\\
 & ... \\ \midrule
Analysis Method & Please follow the format below when answering the questions: \\
(ReAct) & Question: the question you must answer \\
 & Thought: you should always think about what to do \\
 & Action: the tool for information collection, should be one of\\
 &  [Tool Name 1, Tool Name 2, ...]  \textbf{(Contents of Table 2)}\\
 & Action Input: the input to the tool \\
 & Observation: the result of information collection \\
 & ... (You can repeat this Thought/Action/Action Input/Observation \\
 & N times to derive your answer.) \\
 & Thought: I now know the final answer \\
 & Final Answer: the final answer to the original question \\
 & You must derive your final answer based on no more than 10 actions. \\ \midrule
Output Format & After the Final Answer is determined, output the analysis results \\
 & in JSON format according to the following key: \\
 & - result: True or False (result of URL scam determination) \\
 & - scam\_type: Fake online shopping website (specific type of scam) \\
 & - reason: State your decision based on the scam website's features \\ \midrule
Analysis Process & Begin! \\
 & Question: Please analyze this URL https://example.com \\
 & Thought: ...  \\ 
 & Action: ... \\
 & Action Input: ... \\
 & Observation: ... \\
 & ...\textbf{ (Repeat Thought/Action/Action/Input/Observation) }... \\
\bottomrule
\end{tabular}
}
\label{tab:prompt}
\end{table}

We propose ScamFerret, a novel system that addresses the three challenges outlined previously.
Our approach leverages LLMs as autonomous agents to drive the information collection and analysis process, capitalizing on their text comprehension capabilities.
Figure~\ref{fig:proposed_system} provides an overview of the system workflow.

ScamFerret takes a URL as input and proceeds to autonomously collect and analyze relevant information.
The system outputs a classification of whether the website is a scam, the specific type of scam (e.g., fake online shopping), and the rationale for this determination (e.g., non-existent operating company).

The core of ScamFerret's functionality relies on a carefully designed prompt template that guides the interactions between the system and the LLM.
The template in Table~\ref{tab:prompt} is crucial for eliciting appropriate responses from the LLM.
The design of effective prompts has been an active area of research, with several studies exploring techniques to optimize LLM outputs for specific tasks~\cite{DBLP:conf/iclr/ZhouMHPPCB23,DBLP:journals/corr/abs-2302-11382}.
In the following subsections, we describe each component of ScamFerret in detail.

\subsection{\Circled{1} Scam Website Analysis}
ScamFerret analyzes input URLs for potential scams using a multi-step process.
The system first evaluates the URL based on the information embedded in the template (i.e., the target URL for analysis) and the LLM's pre-trained knowledge, following the Task Setting and Analysis Process in Table~\ref{tab:prompt}.
If the initial information is insufficient, ScamFerret performs External Information Collection (\Circled{2}) and re-analyzes the website.
This process iterates until a final determination is made.

\noindent\textbf{Feature Analysis of Scam Websites.}
Research has shown that including specific features and cautions in LLM prompts improves performance~\cite{DBLP:journals/access/KoideNC24}.
We incorporate nine common scam website features into the prompts, leveraging the model's text comprehension abilities (Characteristics Examples, Table~\ref{tab:prompt}).
These features include uncommon pricing, large monetary gifts, language targeting psychological weaknesses, and diverse product/service offerings.
Additionally, we consider lack of company information, inappropriate business contact details, poorly written privacy policies, uncommon payment methods, and outdated information.
This approach differs from traditional machine learning systems that rely on complex, scam-specific feature engineering.
By describing these features in natural language, the LLM can analyze and identify suspicious elements in the collected information.

\noindent\textbf{Tool Selection.}
While LLMs possess extensive knowledge, providing additional external information can enhance their performance.
Studies have shown that external source information significantly improves response accuracy for challenging tasks~\cite{DBLP:journals/corr/abs-2312-10997}.
We define a set of tools that provide useful information for scam website analysis (Tool Definitions, Table~\ref{tab:prompt}).
Each tool includes a name, description, and required input information.
The LLM uses this information to select appropriate tools and extract necessary inputs.
Section~\ref{subsec:tool} details the tools used in this study.

\noindent\textbf{LLM Decision-making.}
ScamFerret utilizes the REasoning and ACTing (ReAct) framework for its decision-making process~\cite{DBLP:conf/iclr/YaoZYDSN023}.
ReAct is an innovative approach that combines reasoning and action, allowing AI systems to articulate their thought processes and adapt their actions based on new information, similar to how humans think and act.
We chose ReAct because it enables LLMs to articulate their reasoning steps, which is crucial for analyzing potential scam websites.
This verbalization of thought processes enhances the model's ability to explain its decision-making rationale.
The decision-making process, implemented using LangChain (an open-source framework for building LLM-based applications that enables the creation of chains of actions for processing tasks)~\cite{langchain}, follows these steps:
\begin{itemize}
\item Repeat the process until the given URL can be identified as a type of website (i.e., scam or legitimate).
\item Scam Website Analysis (\Circled{1}) is performed based on the information embedded in the prompt template.
\item If the LLM determines that there is insufficient information for identification, it will perform External Information Collection (\Circled{2}), embed this information into the template, and then conduct Scam Website Analysis (\Circled{1}) again.
\item If the LLM determines that the information is sufficient for identification, it will perform Analysis Results Output (\Circled{3}) based on the results of all previous analyses.
\end{itemize}

To prevent infinite loops, ScamFerret imposes a limit of 10 tool selections per URL (Analysis Method, Table~\ref{tab:prompt}).
This constraint ensures efficient processing while allowing for thorough investigation.
This iterative approach enables ScamFerret to autonomously gather and analyze information, leading to accurate scam detection.
By combining LLM-based reasoning with strategic tool usage, our system can adapt to various scam scenarios and provide detailed justifications for its conclusions.
Even as the types of prevalent scam websites evolve, ScamFerret can flexibly respond without requiring major updates, as its analytical processes remain universal across various scam scenarios.

\subsection{\Circled{2} External Information Collection}
\label{subsec:tool}
\begin{table}[!t]
\scriptsize
\tabcolsep=0.5mm
\centering
\caption{List of Defined Tools}
{\renewcommand\arraystretch{0.85}
\begin{tabular}{lll} \toprule
Tool Type & Tool Name & Description \\ \midrule
Web & Access URL & A tool that accesses a URL to obtain a status code. \\
Content & & This tool requires a URL as an argument. \\
& Extract Text & A tool that extracts text in the HTML. \\
& & You must access a URL first before using this tool. \\
& & This tool requires the URL as an argument. \\
& Extract Hyperlink & A tool that extracts a-tag hyperlinks and texts in the HTML.\\
& & You must access a URL first before using this tool. \\
& & This tool requires the URL as an argument. \\ \midrule
Search & Get Search Result & A tool to retrieve search results from a search engine.\\
Engine & & This tool requires a search query as an argument. \\
& & You cannot use a URL as-is as a search query. \\
& & Note that only the top 10 results will be retrieved. \\ \midrule
Social & Search X/Twitter & A tool to retrieve posts containing a keyword from X/Twitter. \\
Media & & This tool requires a search query as an argument. \\
& & You cannot use a URL as-is as a search query. \\
& & Note that only the latest top 10 results will be retrieved. \\
& Search Reddit & A tool to retrieve posts containing a keyword from Reddit. \\
& & This tool requires a search query as an argument. \\
& & You cannot use a URL as-is as a search query. \\
& & Note that only the top five related posts and the top five \\
& & associated comments will be retrieved.\\ \midrule
WHOIS & Retrieve WHOIS & A tool to retrieve domain name information from WHOIS. \\
& & This tool requires a domain name as an argument. \\ \midrule
DNS & Retrieve DNS Record & A tool to retrieve DNS records using the dig command. \\
Lookup & & This tool requires a domain name as an argument. \\ \midrule
TLS  & Retrieve Certificate & A tool to retrieve certificate information from crt.sh. \\
Certificate & & This tool requires a domain name as an argument. \\
& & Note that only the latest top 5 results will be retrieved. \\
\bottomrule
\end{tabular}
}
\label{tab:tools}
\end{table}

ScamFerret collects external information using tools selected during the Scam Website Analysis (\Circled{1}) phase.
We designed these tools to capture the inherent characteristics of scams that attempt to deceive users, rather than focusing on specific scam types.
Our approach is informed by previous studies on scam websites.
The tools collect information from external sources that are likely to yield traces indicative of scam websites, enabling LLMs to analyze sophisticated scams effectively.

ScamFerret allows the LLM to determine which tools to use for information collection, meaning that not all tools are used in every analysis, and some may be used only once.
The system may also select the same tool multiple times when needed, such as when analyzing multiple domain names found in the collected information.
We have defined six categories encompassing a total of nine tools, as shown in Table~\ref{tab:tools}, to collect information from various perspectives.
These tools can obtain information commonly considered by human analysts when analyzing scam websites and that has been reported to be effective for detection in previous studies~\cite{DBLP:conf/sp/BitaabCOLWAWBSD23,DBLP:conf/acsac/KotziasRPSB23,DBLP:conf/ndss/LiYN23}.

\noindent\textbf{Web Content.}
Attackers often use text, images, and other web content to deceive users.
We analyze these elements using three tools built with Playwright~\cite{playwright}, which can render JavaScript and interact dynamically with websites.
The Access URL tool takes a URL as input and retrieves the HTTP response status code.
The Extract Text tool extracts strings (i.e., the innerText of HTMLElements) from the HTML content of the page accessed by the Access URL tool.
To avoid including irrelevant strings, we target a maximum of three HTML tags in the same hierarchy.
The Extract Hyperlink tool extracts combinations of the \textit{href} attribute and the text within \textit{<a>} tags (e.g., (http://example.com/contact.html, Contact Page)) from the HTML content of the page accessed by the Access URL tool.
To ensure relevance, we extract text content from the same level as the \textit{<a>} tag and one level below it in the HTML DOM tree structure.

These tools can be used recursively for detailed content analysis.
For instance, after accessing the top page, if company information is not present in the extracted text, the hyperlinks can be analyzed to locate and access a dedicated company information page.
While it's possible that the HTML of scam websites is obfuscated, these tools are not affected because they analyze the displayed character strings and \textit{<a>} tag elements to extract information.

\noindent\textbf{Search Engine.}
Search engines provide valuable information about user-reported scam websites.
We implemented the Get Search Result tool using Tavily~\cite{tavily}, a commercial search engine API designed for LLMs that provides search results in an LLM-interpretable format.
As illustrated in Figure~\ref{fig:proposed_system}, this tool can collect information such as the reputation of a company operating the website under analysis.
The tool takes a search query as input and returns relevant URLs and web page content summaries, enabling comprehensive assessment of potential scam websites.

\noindent\textbf{Social Media.}
Social media platforms can offer security-related information posted by various users~\cite{DBLP:conf/sp/BitaabCOLWAWBSD23,DBLP:conf/ccs/TangML0022}.
The real-time nature of social media allows for quick gathering of scam reports and website reputations, often before they appear on dedicated review websites.
We created two tools to retrieve posts by keyword search.
The Search X/Twitter tool uses the X/Twitter API~\cite{twitter_api} to retrieve up to 10 latest posts containing the specified keyword.
The Search Reddit tool uses the Reddit API~\cite{reddit_api} to retrieve up to 5 related posts and 5 associated comments with the specified keyword.

\noindent\textbf{WHOIS Information.}
Attackers often launch websites shortly after acquiring domain names~\cite{DBLP:conf/uss/OestZWNBZTDA20}, resulting in recently registered domains.
In contrast, legitimate websites typically have longer operational histories and well-defined management information.
To analyze these characteristics, we implemented the Retrieve WHOIS tool.
This tool uses the Linux Shell Command ``whois'' to retrieve information including the domain registrant, registration date, administrator, and managing organization.

\noindent\textbf{DNS Record.}
DNS records can provide useful information for distinguishing between scam and legitimate websites.
For example, NS and SOA record settings may differ significantly between scam and legitimate operations~\cite{DBLP:conf/imc/HaoFP11}.
We implemented the Retrieve DNS Record tool, which uses the Linux Shell Command ``dig [record type] [domain name] @8.8.8.8'' to obtain DNS records such as A, AAAA, NS, SOA, TXT, and MX.

\noindent\textbf{TLS Certificate.}
Modern scam websites often use TLS certificates, with potential biases toward specific certification authorities~\cite{DBLP:conf/asiaccs/KimCKDSAD21}.
For instance, attackers may favor free certificates from Let's Encrypt or use the Subject Alternative Name to link multiple domain names to a single certificate.
We implemented the Retrieve Certificate tool, which uses crt.sh~\cite{crtsh} to search the Certificate Transparency log and retrieve a list of certificates associated with a given domain name.

\subsection{\Circled{3} Analysis Results Output}
After iteratively performing Scam Website Analysis (\Circled{1}) and External Information Collection (\Circled{2}), ScamFerret generates a final output based on the LLM's determination of whether the analyzed URL represents a scam or legitimate website.
The analysis results comprise three key components:

\noindent\textbf{Result.} A binary classification indicating whether the URL is associated with a scam (``True'') or a legitimate website (``False'').

\noindent\textbf{Scam Type.} If the URL is classified as a scam, this field specifies the particular category or method of scam detected.

\noindent\textbf{Reason.} A detailed explanation of the LLM's decision-making process, outlining the key factors and evidence that led to the final determination.

These components are generated as part of Output Format in Table~\ref{tab:prompt}.
The structured output allows for clear interpretation of the analysis results, providing both a concise classification and the underlying rationale.
This approach enhances the transparency and interpretability of the system's decision-making process, which is crucial for both end-users and further research in the field of online scam detection.

\section{Dataset}
\label{sec:dataset}
\begin{table}[!t]
\tabcolsep=1.5mm
  \centering
  \notsotiny
    \caption{Ground-truth Dataset for Evaluation}
    {\renewcommand\arraystretch{0.85}
    \begin{tabular}{llrr} \toprule
      Scam Type & Language & \# of Scam Websites & \# of Legitimate Websites \\ \midrule
      Online Shopping & English & 200 & 200 \\ 
       & Japanese & 200 & 200 \\ 
       & German & 200 & 200 \\ \midrule
       Technical Support & English & 200 & 200 \\ \midrule
       Cryptocurrency & English & 200 & 200 \\ \midrule
       Investment & English & 200 & 200 \\ \midrule
       Total & 3 Languages & 1,200 & 1,200 \\ 
       \bottomrule
    \end{tabular}
    }
  \label{tab:dataset}
\end{table}

To assess ScamFerret's accuracy in detecting challenging scam websites, we created a new ground-truth dataset with verified labels.
Existing public datasets~\cite{eng_beyond,eng_cryptoscam} were mostly inaccessible or unverifiable for our evaluation. Our dataset creation process involved collecting candidates, then establishing a reproducible ground-truth through four main steps.

\subsection{Candidate Collection}
Due to the challenges in detecting modern scam websites using traditional anti-virus engines and services like VirusTotal~\cite{virustotal}, we created a custom dataset for evaluation.
Our dataset comprises four types of English scam websites: Fake Online Shopping, Technical Support Scam, Cryptocurrency Scam, and Investment Scam.
We collected these from five up-to-date public sources between April 1 and April 7, 2024~\cite{eng_onlineshopping_fraud1,eng_onlineshopping_fraud2,eng_cryptoscam,eng_tss_fraud1,eng_tss_fraud2}.
For Fake Online Shopping, we also included scam websites in German and Japanese, which were collected during the same period from two additional disclosure websites~\cite{jpn_onlineshopping_fraud,ger_onlineshopping_fraud}.

To assess ScamFerret's performance accurately, we compiled a corresponding dataset of legitimate websites.
Unlike phishing websites, the scam websites in our study lack direct legitimate counterparts.
We aimed to collect diverse legitimate websites to demonstrate that ScamFerret's classification is not based solely on website strings or domain names containing words like ``shopping'' or ``support''.

We utilized Curlie~\cite{curlie} and Trustpilot~\cite{trustpilot} to create our legitimate website dataset.
Curlie, a manually compiled web directory, organizes multilingual websites into categories.
Trustpilot is a user-driven review platform for business services and products.
These sources have been effectively used in previous studies for domain name classification and creating legitimate website datasets~\cite{DBLP:conf/imc/VallinaPFPGBHTV20,DBLP:conf/acsac/KotziasRPSB23}.
From Curlie and Trustpilot, we collected information for four types of \textit{legitimate} websites (Online Shopping, Technical Support, Cryptocurrency, and Investment) in English.
For \textit{legitimate} Online Shopping, we collected data in English, German, and Japanese.

\subsection{Ground-truth Dataset Creation}
We create a ground-truth dataset of scam and legitimate websites through a four-step selection process from collected candidates.

\noindent\textbf{Top List Filtering.}
We excluded websites listed in the top 100,000 domain names of the Tranco List~\cite{DBLP:conf/ndss/PochatGTKJ19}, a widely used reference for legitimate websites in research.
This step helped eliminate obviously legitimate websites that did not need to be analyzed in the LLM from the analysis target.
Note that while we gathered data from Curlie and Trustpilot, most of the websites collected were minor sites, resulting in few that ranked within the top 100,000.

\noindent\textbf{URL Accessibility Check.}
To detect active scam websites in real-time with high accuracy, we first excluded inaccessible URLs.
We used Playwright to simulate common user access with a standard user agent (i.e., Mozilla/5.0 (Windows NT 10.0; Win64; x64) AppleWebKit/537.36 (KHTML, like Gecko) Chrome/122.0\\.0 Safari/537.36).
URLs that did not return an HTTP status code of 200 were excluded from the dataset.

\noindent\textbf{Manual Inspection.}
We manually verified each URL's appropriateness for our study and its proper categorization.
Three security engineers examined each URL using search engines and analyzed the web content and screenshots.
Due to individual language limitations, the evaluators collaborated to reach a consensus, excluding URLs that did not match the specified scam type.

\noindent\textbf{Random Sampling.}
To create a balanced dataset, we randomly sampled an equal number of scam and legitimate websites for each type and language.
This approach ensures an accurate evaluation of ScamFerret's classification performance.
The final dataset comprises 1,200 scam websites and 1,200 legitimate websites, with 200 URLs for each language and type combination, as shown in Table~\ref{tab:dataset}.

\section{Evaluation}
\label{sec:evaluation}
We evaluate both the classification accuracy and explanation quality of ScamFerret using the ground-truth dataset described in Section~\ref{sec:dataset}.

\subsection{Experimental Setup}
\noindent\textbf{Models.}
We compare the performance of three LLMs: GPT-3.5 (gpt-35-turbo-1106) and GPT-4 (gpt-4-1106-preview) from OpenAI, accessed via Azure OpenAI Service~\cite{azure_openai}, and Gemini (Gemini 1.5 Pro) from Google DeepMind~\cite{gemini}.
While LLMs have content filters to protect against harmful content and other issues, we disabled these filters for GPT-3.5 and GPT-4 in our experiments.
It is important to note that these model performances are as of the experiments conducted in April 2024, and the models may have been updated since then.

\noindent\textbf{Parameters.}
For each model, we configured two key parameters: the context size, which limits the input text length, and the temperature, which controls output diversity.
Preliminary experiments showed that ScamFerret performed optimally with a context size of 128,000 (the maximum allowed) and a temperature of 0.7, which was found to be the most effective among tested values.
We applied these settings in our main experiments.

\noindent\textbf{Conventional Systems.}
We evaluate ScamFerret against three conventional systems:

\noindent\textbf{\textit{{Single-turn Prompt}}}: A system that uses a brief prompt to query LLMs for scam detection.
This prompt includes the role of a scam detection expert, features of scam websites, web content, and specifies a JSON output format.
Unlike ScamFerret, it analyzes the website only once.

\noindent\textbf{\textit{{Beyond Phish~\cite{DBLP:conf/sp/BitaabCOLWAWBSD23}}}}: A binary classification system for online shopping scams, with publicly available implementation~\cite{eng_beyond}.

\noindent\textbf{\textit{Scamdog Millionaire~\cite{DBLP:conf/acsac/KotziasRPSB23}}}: A supervised learning approach for binary classification of online shopping scam websites based on extracted web content features.

For the conventional systems, we created new training datasets to replicate their functionality, as the original training data was not available.
We used 1,600 English websites (200 each for scam and legitimate across four types) that were not included in our ground-truth dataset.
Features were generated based on information from the original papers, and models were trained to classify websites as scam or legitimate using URLs as input.

We evaluated the Single-turn Prompt on all scam websites in our ground-truth dataset, leveraging its multilingual capabilities.
The two conventional systems, designed for English websites, were evaluated only on the English subset of our dataset.

\subsection{Scam and Legitimate Website Classification Accuracy}
\label{subsec:evaluation_accuracy}
We evaluate the binary classification performance (scam vs. legitimate) separately for each combination of scam type and language in our ground-truth dataset using standard metrics.

\begin{table}[!t]
\vspace{-1em}
\tabcolsep=1.0mm  
\centering  
\notsotiny
\caption{Summary of Binary Classification Results}
\begin{tabular}{ll|rrr|rrr}
\toprule
 & & \multicolumn{3}{c|}{ScamFerret} & \multicolumn{3}{c}{Single-turn Prompt} \\
 & & \multicolumn{3}{c|}{(Proposed System)} &\multicolumn{3}{c}{}  \\
  
 & & GPT-4 & GPT-3.5 & Gemini & GPT-4 & GPT-3.5 & Gemini \\
\midrule
\multirow{4}{*}{\begin{tabular}{@{}l@{}}\\Overall Results for\\Four Scam Types\\in English\end{tabular}} 
 & Accuracy & \textbf{0.972} & 0.938 & 0.887 & 0.833 & 0.803 & 0.781 \\
 & TPR/Recall & \textbf{0.964} & 0.913 & 0.848 & 0.790 & 0.786 & 0.676 \\
 & TNR & \textbf{0.980} & 0.964 & 0.926 & 0.875 & 0.820 & 0.886 \\
 & Precision & \textbf{0.980} & 0.962 & 0.920 & 0.863 & 0.814 & 0.856 \\
 & F1 score & \textbf{0.972} & 0.936 & 0.882 & 0.825 & 0.800 & 0.755 \\
\midrule
\multirow{4}{*}{\begin{tabular}{@{}l@{}}\\Overall Results for\\Online Shopping Websites\\in Three Languages\end{tabular}} 
 & Accuracy & \textbf{0.993} & 0.928 & 0.892 & 0.891 & 0.872 & 0.811 \\
 & TPR/Recall & \textbf{0.988} & 0.872 & 0.840 & 0.847 & 0.858 & 0.688 \\
 & TNR & \textbf{0.997} & 0.985 & 0.943 & 0.935 & 0.885 & 0.933 \\
 & Precision & \textbf{0.997} & 0.983 & 0.937 & 0.929 & 0.882 & 0.912 \\
 & F1 score & \textbf{0.992} & 0.924 & 0.886 & 0.886 & 0.870 & 0.784 \\
\bottomrule  
\end{tabular}
\label{tab:results_evaluation_summary}
\vspace{-1em}
\end{table}

\begin{table}[!t]
\vspace{-1em}
\tabcolsep=1.0mm  
\centering  
\notsotiny
\caption{Binary Classification Results for English Online Shopping Websites}
\begin{tabular}{l|rrr|rrr|r|r}
\toprule
 & \multicolumn{3}{c|}{ScamFerret} & \multicolumn{3}{c|}{Single-turn Prompt} & \multicolumn{1}{c|}{Beyond} & \multicolumn{1}{c}{Scamdog}\\
  & \multicolumn{3}{c|}{(Proposed System)} &\multicolumn{3}{c|}{} & \multicolumn{1}{c|}{Phish~\cite{DBLP:conf/sp/BitaabCOLWAWBSD23}} & \multicolumn{1}{c}{ Millionaire~\cite{DBLP:conf/acsac/KotziasRPSB23}} \\
  
 & GPT-4 & GPT-3.5 & Gemini & GPT-4 & GPT-3.5 & Gemini & - & -\\
\midrule
Accuracy & \textbf{0.993} & 0.923 & 0.870 & 0.873 & 0.863 & 0.790 & 0.883 & 0.915 \\
TPR/Recall & \textbf{0.985} & 0.880 & 0.820 & 0.805 & 0.825 & 0.625 & 0.815 & 0.890 \\
TNR & \textbf{1.000} & 0.965 & 0.920 & 0.940 & 0.900 & 0.955 & 0.950 & 0.940 \\
Precision & \textbf{1.000} & 0.962 & 0.911 & 0.931 & 0.892 & 0.933 & 0.942 & 0.937 \\
F1 score & \textbf{0.992} & 0.919 & 0.863 & 0.863 & 0.857 & 0.749 & 0.874 & 0.913 \\
\bottomrule  
\end{tabular}
\label{tab:results_evaluation_shopping_english}
\vspace{-1em}
\end{table}

\begin{table}[!t]
\vspace{-1em}
\tabcolsep=1.0mm  
\centering  
\notsotiny
\caption{Multi-class Classification Results}
\begin{tabular}{ll|rrr|rrr}
\toprule
 & & \multicolumn{3}{c|}{ScamFerret} & \multicolumn{3}{c}{Single-turn Prompt} \\
 & & \multicolumn{3}{c|}{(Proposed System)} &\multicolumn{3}{c}{} \\
  
 & & GPT-4 & GPT-3.5 & Gemini & GPT-4 & GPT-3.5 & Gemini \\
\midrule
Overall Results for & TPR/Recall & \textbf{0.860} & 0.664 & 0.240 & 0.679 & 0.538 & 0.158 \\
Four Scam Types & Precision & \textbf{0.977} & 0.948 & 0.765 & 0.845 & 0.750 & 0.581 \\
in English & F1 score & \textbf{0.915} & 0.781 & 0.365 & 0.753 & 0.627 & 0.248 \\
\midrule 
Overall Results for & TPR/Recall & \textbf{0.982} & 0.767 & 0.830 & 0.827 & 0.858 & 0.633 \\
Online Shopping Websites & Precision & \textbf{0.997} & 0.981 & 0.927 & 0.929 & 0.882 & 0.905 \\
in Three Languages & F1 score & \textbf{0.989} & 0.861 & 0.880 & 0.874 & 0.870 & 0.745 \\
\bottomrule  
\end{tabular}
\label{tab:results_multicalss}
\vspace{-1em}
\end{table}

\noindent\textbf{Evaluation Metrics.}
We use four main classification outcomes:

\noindent\textbf{\textit{True Positive (TP):}}
Correctly identified scam website.

\noindent\textbf{\textit{True Negative (TN):}}
Correctly identified legitimate website.

\noindent\textbf{\textit{False Positive (FP):}}
Legitimate website misclassified as scam.

\noindent\textbf{\textit{False Negative (FN):}}
Scam website misclassified as legitimate.

From these, we derive the following performance metrics:

\noindent\textbf{\textit{Accuracy:}}
The overall correct classification rate, calculated as $Accuracy = \frac{TP + TN}{TP + TN + FP + FN}$.

\noindent\textbf{\textit{True Positive Rate (TPR) / Recall:}}
The proportion of correctly identified scam websites, given by 
$TPR/Recall = \frac{TP}{TP + FN}$.

\noindent\textbf{\textit{True Negative Rate (TNR):}}
The proportion of correctly identified legitimate websites, expressed as $TNR = \frac{TN}{TN + FP}$.

\noindent\textbf{\textit{Precision:}}
The proportion of correct scam identifications among all identified scam websites, defined as $Precision = \frac{TP}{TP + FP}$.

\noindent\textbf{\textit{F1 score:}
The harmonic mean of precision and recall, providing a balanced measure of the system's performance, calculated as $F1 score = \frac{2 \times Precision \times Recall}{Precision + Recall}$.}

Our evaluation employs two classification methods.
For binary classification, the system's performance is evaluated based on the \textit{Result} field output (``True'' or ``False'').
For multi-class classification, the evaluation considers the system's \textit{Scam Type} field, where semantically equivalent responses are considered correct (e.g., both ``Fake investment site'' and ``Fake financial services site'' are accepted for Investment scams).
The evaluation uses TPR/Recall, Precision and F1 score as performance metrics, since our system only outputs the scam type when it classifies a URL as a scam website.
With an equal distribution of URLs across scam types, we employ macro-averaging for evaluation.
These metrics collectively provide a comprehensive assessment of the classifier's performance in distinguishing between scam and legitimate websites.

\noindent\textbf{Summary of Binary Classification Results.}
Table~\ref{tab:results_evaluation_summary} presents the classification accuracy comparison between ScamFerret and the Single-turn Prompt.
ScamFerret (GPT-4) achieved mean scores of 0.972 (Accuracy), 0.964 (TPR), 0.980 (TNR), 0.980 (Precision), and 0.972 (F1 score) across four English scam types: Online Shopping, Technical Support, Cryptocurrency, and Investment.
The results demonstrate that ScamFerret's multi-tool information gathering approach significantly outperforms the Single-turn Prompt method using only URLs and top page content, with TPR improving from 0.790 to 0.964 and TNR from 0.875 to 0.980.

For Online Shopping websites in English, German, and Japanese, ScamFerret (GPT-4) demonstrated robust performance with mean scores of 0.993 (Accuracy), 0.988 (TPR), 0.997 (TNR), 0.997 (Precision), and 0.992 (F1 score).
While all models performed well in identifying legitimate Online Shopping websites (TNR > 0.940), ScamFerret's external information integration significantly improved TPR across the three languages (0.988 vs. 0.847 for Single-turn Prompt), confirming its effectiveness in classifying Online Shopping websites regardless of language.
Compared to GPT-3.5 and Gemini, GPT-4 significantly improved TPR for German (0.990 vs. 0.810 and 0.790) and Japanese (0.990 vs. 0.925 and 0.910).
These findings suggest that advanced LLMs can effectively perform expert-level analysis across multiple languages, potentially eliminating the need for language-specific expertise in scam detection.

\noindent\textbf{Analysis of False Positives (FPs).}
We analyzed 18 false positive cases in ScamFerret using the GPT-4 model, resulting in an overall false positive rate of 1.5\%.
These cases were primarily in Cryptocurrency (16) and Online Shopping (German) (2).
Three main characteristics were identified:

\noindent\textbf{\textit{User-inciting phrases}}:
15 websites contained phrases like ``free shipping'' or ``unusually large financial returns,'' which are common in both legitimate and scam websites, making it challenging for LLMs to differentiate.

\noindent\textbf{\textit{Negative reviews}}:
A high number of negative posts on review websites, even for legitimate websites, led to false positives.

\noindent\textbf{\textit{Privacy protection services}}:
The use of WHOIS privacy protection was considered suspicious, despite being a common practice for both legitimate and scam websites.

To address these issues, including context about these features in the prompts could improve classification accuracy.

\noindent\textbf{Analysis of False Negatives (FNs).}
We analyzed false negative cases in ScamFerret using the GPT-4 model, identifying 33 instances across various categories: Online Shopping (English) (3), Technical Support (10), Cryptocurrency (14), Investment (2), Online Shopping (German) (2), and Online Shopping (Japanese) (2). The overall false negative rate was 2.75\%.
Three main characteristics were observed:

\noindent\textbf{\textit{Domain status changes}}:
All 14 Cryptocurrency cases were related to subdomains of \textit{stockfund.co}, which had a \textit{pendingDelete} status during the experiment.
This led to inconsistent LLM analysis results for related domain names, highlighting the need for caution in operational settings where LLMs may produce split judgments in ambiguous situations.

\noindent\textbf{\textit{Inconclusive evidence}}:
16 website judgments were based on suspicious aspects of website content and WHOIS information, but lacked definitive evidence.
This suggests a need to enhance the tool to collect more relevant information for analysis.

\noindent\textbf{\textit{Lack of review information}}:
The absence of relevant URLs on review websites for reporting spam led to incorrect judgments.
The LLM interpreted this lack of information as an indicator of a legitimate website.
This highlights the need to improve prompts by considering that the absence of information should not be used as a sole basis for classification.

These findings indicate areas for improvement in both the information collection process and the LLM's decision-making capabilities.
Future work should focus on refining the prompts to account for these scenarios and enhancing the tool's ability to gather more comprehensive and relevant data for analysis.

\noindent\textbf{Comparison of Results with Conventional Systems.}
As shown in Table~\ref{tab:results_evaluation_shopping_english}, when comparing the results for English Online Shopping websites, ScamFerret demonstrated superior performance compared to two conventional systems (Beyond Phish and Scamdog Millionaire).
While the conventional systems achieved 0.883 and 0.915 accuracy rates through machine learning on structurally similar website datasets, ScamFerret outperformed them using LLM capabilities without requiring any additional training.
These results suggest that the extensive knowledge base of LLMs provides a significant advantage over conventional machine learning models in classifying scam and legitimate Online Shopping websites.

\noindent\textbf{Multi-class Classification Results.}
Table~\ref{tab:results_multicalss} presents the macro-averaged results of multi-class classification across Online Shopping (English, German, Japanese), Technical Support, Cryptocurrency, and Investment by scam type and language.

ScamFerret with GPT-4 achieved the highest performance with TPR/Recall of 0.860, Precision of 0.977, and F1 score of 0.915.
Compared to binary classification (Table~\ref{tab:results_evaluation_summary}), TPR/Recall decreased from 0.964 to 0.860, mainly due to 9 Cryptocurrency and 24 Investment scams being misclassified as fake shopping sites.
Gemini's TPR/Recall dropped significantly from 0.848 to 0.240, as it misclassified most Cryptocurrency and Investment scams as fake shopping sites.
These results indicate that even GPT-4 struggles with accurate multi-class scam categorization.

For three-language Online Shopping classification, ScamFerret with GPT-4 maintained high performance (TPR/Recall: 0.982, Precision: 0.997, F1: 0.989).
GPT-3.5 and Gemini also retained accuracy levels similar to their binary classification results, demonstrating effective fake shopping site detection across languages.
The Single-Turn Prompt approach performed consistently lower than binary classification and failed to match ScamFerret's performance across all categories, confirming ScamFerret with GPT-4's superiority in multi-class classification.

\subsection{Information Used for Website Analysis}
We conducted a detailed analysis of the tools employed by the LLM and the key characteristics cited in its decision-making process.

\begin{table}[!t]
 \tabcolsep=1.0mm
  \centering
  \notsotiny
    \caption{Number of Tools Selected and Usage per LLM}
    {\renewcommand\arraystretch{0.85}
        \begin{tabular}{l|rr|rr|rr} \toprule
         Tool&\multicolumn{2}{c|}{ScamFerret (GPT-4)}& \multicolumn{2}{c|}{ScamFerret (GPT-3.5)}& \multicolumn{2}{c}{ScamFerret (Gemini)}\\
        & \# Selected & \# Used& \# Selected& \# Used & \# Selected & \# Used\\ \midrule
        Access URL &2,724&100\%& 2,417& 96.7\%& 2,471& 77.8\%\\ 
        Extract Text&2,723&99.0\%& 2,398& 95.3\%& 3,406& 85.7\%\\ 
        Extract Hyperlink&1,018&40.5\%& 281& 11.3\%& 1,088& 34.6\%\\ 
        Get Search Result&1,798&67.6\%& 51& 2.13\%& 552& 17.5\%\\ 
        Search X/Twitter&1,060&43.5\%& 22& 0.88\%& 270& 9.83\%\\ 
        Search Reddit&1,545&63.5\%& 12& 0.50\%& 196& 7.12\%\\ 
        Retrieve WHOIS&2,479&99.5\%& 1,128& 45.9\%& 419& 16.3\%\\ 
        Retrieve DNS Record&617&25.5\%& 51& 2.13\%& 129& 5.04\%\\         
        Retrieve Certificate&1,276&52.8\%& 85& 3.54\%& 83& 3.33\%\\
        \bottomrule
    \end{tabular}
    }
    \label{tab:tools_results}
\end{table}

\begin{table}[!t]
\tabcolsep=1.5mm
  \centering
  \notsotiny
    \caption{Selected Information Types and Keywords}
    {\renewcommand\arraystretch{0.85}
    \begin{tabular}{l|l} \toprule
     Information Type & Keywords \\
     \midrule
      Certificate Information & TLS, certificate, HTTPS, SSL \\ \midrule
      Company Information & company information, non-existent companies,\\
       & non-existent company, physical address \\ \midrule
      Contact Information & email, phone number, contact information, toll-free number \\ \midrule
      Domain Name & WHOIS, registrant, privacy service, domain, DNS \\ \midrule
      Payment Method & payment, Bitcoin, cryptocurrency \\ \midrule
      Privacy Information & privacy policy, privacy notation, privacy policies \\ \midrule
      Social Engineering & psychological, lure, urgency, unrealistic, phishing tactic, \\
       & scam tactic, short timeframe \\ \midrule
      Unusual Price & abnormal price, low price, discounts, free items, \\
       & high return, guaranteed returns, free delivery, free shipping \\ \midrule
      User Review & social media, feedback, review, Twitter, Reddit, complaint, \\
       & report, discussion, forum, low trust score, negative, \\
       & indicators, social platforms \\ \midrule
       Website Status & update, copyright, outdated, up-to-date \\
      \bottomrule
    \end{tabular}
    }
    \label{tab:list_keywords}
\end{table}

\begin{table}[!t]
 \tabcolsep=1.0mm
  \centering
  \notsotiny
    \caption{Information in Reasons for Website Decision}
    {\renewcommand\arraystretch{0.85}
        \begin{tabular}{l|r|r|r} \toprule
         Information&ScamFerret (GPT-4)& ScamFerret (GPT-3.5) & ScamFerret (Gemini) \\
        & \# Reasons & \# Reasons & \# Reasons \\ \midrule
        Certificate Information & 770 (32.1\%)& 43 (1.80\%)& 28 (1.17\%)\\ 
        Company Information & 307 (12.8\%)& 405 (16.9\%)& 87 (3.62\%)\\ 
        Contact Information & 621 (25.9\%)& 227 (9.46\%)& 233 (9.71\%)\\ 
        Domain Name & 1,866 (77.8\%)& 952 (39.7\%)& 157 (6.54\%)\\ 
        Payment Method & 339 (14.1\%)& 315 (13.1\%)& 81 (3.38\%)\\ 
        Privacy Information & 379 (15.8\%)& 229 (9.54\%)& 45 (1.88\%)\\ 
        Social Engineering & 796 (33.2\%)& 279 (11.6\%)& 108 (4.50\%)\\ 
        Unusual Price& 1,104 (46.0\%)& 686 (28.6\%)& 501 (20.9\%)\\ 
        User Review & 1,544 (64.3\%)& 52 (2.17\%)& 90 (3.75\%)\\ 
        Website Status & 294 (12.3\%)& 165 (6.88\%)& 69 (2.88\%)\\ 
        \bottomrule
    \end{tabular}
    }
    \label{tab:llm_reason}
\end{table}

\noindent\textbf{Tools Used for Website Analysis.}
We analyzed the tools selected by the LLM to evaluate scam and legitimate websites.
Table~\ref{tab:tools_results} shows the number of times each tool was selected (\# Selected) and the percentage of the entire dataset in which the tool was used (\# Used) for GPT-4, GPT-3.5, and Gemini.
Tool selection and utilization varied significantly across models.
Please note that the total number of tool selections may exceed the dataset maximum of 2,400, as the same tool can be chosen multiple times within a single analysis.

For GPT-4, the most frequently used tools were Access URL (100\%), Retrieve WHOIS (99.5\%), and Extract Text (99.0\%).
GPT-4 demonstrated sophisticated tool combinations, such as accessing a given URL and extracting web content text upon confirming a 200 OK HTTP status.
It also showed the ability to autonomously collect necessary information by recursively using Access URL and Extract Text when encountering relevant strings (e.g., ``about'', ``privacy policy'', ``payment'') on the top page.
The model effectively used search engine and social media search tools to analyze websites based on external information, searching for ``[domain name] review'' and ``[extracted company name]''.
GPT-4 demonstrated human-like analytical capabilities by selecting and combining various tools as needed, even when there was insufficient information for making a decision (e.g., information could not be obtained using a specific tool).
In contrast, GPT-3.5 and Gemini were limited to using Access URL and Extract Text, unable to fully utilize other tools.
This suggests that a certain level of text comprehension ability is necessary for effective tool selection in analyzing scam websites.

Future work could include tools for analyzing feature similarity to identify scam websites deployed by the same attacker.
This would enhance the system's ability to detect coordinated scams.

\noindent\textbf{Characteristics Included in Decision Basis.}
We analyzed the decisive factors used by ScamFerret in determining scam websites by examining the entire decision basis.
We manually analyzed feature and keyword pairs that were decisive in the reasoning for 120 URLs (10 URLs per type and language from Table~\ref{tab:dataset}) correctly classified by ScamFerret using GPT-4.
Table~\ref{tab:list_keywords} presents the results, showing 47 keywords used across 10 information types.
We then investigated the frequency of these 47 keywords in the overall basis for website judgments across the entire ground-truth dataset.

Table~\ref{tab:llm_reason} shows the information types and their frequency in the website decision rationale for each LLM.
The GPT-4 model, with its wide range of tool selection for acquiring external information, provided judgments from multiple perspectives.
For Domain Name, which was the most common information type, the model often cited characteristics such as ``suspicious due to recent domain registration'' based on WHOIS information.
User Review was the second most common, where the model assessed domain reputation using search engines and social media to incorporate public opinion.
In the Unusual Price and Social Engineering categories, the model appropriately analyzed and identified statements targeting human psychological vulnerabilities or offering unusually inexpensive products or services.

This analysis demonstrates that referencing a wide range of external information enables multi-faceted judgment of website legitimacy, leading to improved detection accuracy and clearer explanations for the decision basis.

\subsection{LLM Cost Analysis}
\noindent\textbf{API Usage Fees.}
We analyzed the cost per URL for ScamFerret using the 2,400 URLs in the Ground-truth Dataset.
Using Azure OpenAI services, the total cost was \$497.39 for GPT-4 (\$0.207 per URL) and \$138.89 for GPT-3.5 (\$0.058 per URL).
Gemini was available free of charge during the experiment (April 2024).
However, if we calculate costs based on current prices (August 2024), Gemini would cost \$402.10 total, or \$0.168 per URL.
It's noteworthy that OpenAI released GPT-4o on August 6, reducing the token cost to one-fourth of GPT-4's previous cost (\$0.01/1k tokens to \$0.0025/1k tokens).
This trend suggests that as LLMs continue to develop, usage costs are likely to decrease further, potentially addressing current cost concerns in the near future.

\noindent\textbf{Execution Time.}
We also analyzed the execution time of ScamFerret in the evaluation experiment in Section~\ref{sec:evaluation} using the 2,400 URLs in the ground-truth dataset.
The total execution time when using GPT-4 was 48 hours, 28 minutes, 4 seconds (1 minute, 13 seconds per URL), GPT-3.5 was 7 hours, 57 minutes, 29 seconds (12 seconds per URL), and Gemini was 29 hours, 58 minutes, 13 seconds (45 seconds per URL).
The execution time was divided between the interaction with the LLM and information retrieval by the tools.
As a result, 79.2\% of the total execution time for GPT-4 was spent interacting with LLM, 33.6\% for GPT-3.5, and 80.7\% for Gemini.
In the current situation, the bottleneck in the analysis of scam websites is the time required for communication with LLM (excluding GPT-3.5, which has a high inference speed).
In particular, GPT-4 achieved excellent results in terms of classification accuracy, but it takes longer to make inferences than other models, which is a major issue for practical application.
However, the development of LLMs has been remarkable, and we believe that this problem will be solved by newly developed models (e.g., GPT-4o and Claude 3.5 Sonnet).

\section{Discussion}
\subsection{LLM-Based Scam Website Detection}
LLMs offer significant advantages over conventional machine learning approaches for scam website detection.
They excel in understanding complex linguistic patterns and contextual nuances, identifying sophisticated scam tactics.
Their adaptability allows effective detection of evolving scams across languages without extensive retraining.
LLMs can autonomously utilize external information collection tools, iteratively gathering and analyzing data when lacking clear scam indicators.
This self-directed process enables more accurate determinations about website legitimacy in challenging cases.
LLMs provide human-interpretable explanations, improving system transparency and reliability.
This approach increases detection accuracy, offers insights into emerging scam patterns, and potentially reduces false positives and costs associated with maintaining conventional machine learning models specialized for detection of each scam type.
While LLMs offer many benefits, it is crucial to be mindful that their probabilistic nature may lead to inconsistent classifications when dealing with sophisticated scam websites that are difficult to identify at first glance.

\subsection{Limitations}
This study presents three primary limitations:

\noindent\textbf{Cost implications of multiple LLM uses.}
The repeated use of LLMs for thought processes increases token generation, significantly raising operational costs.
While services like Azure OpenAI base their pricing on token usage, the extensive tool utilization in our proposed system may lead to higher-than-anticipated expenses in real-world applications.
Mitigation strategies include designing tools for efficient token usage, carefully selecting URLs for analysis, and potentially employing locally executable LLMs like Llama3 for specific analysis targets and languages.

\noindent\textbf{Detection Evasion by Attackers.}
Modern attackers employ sophisticated techniques to evade detection systems by manipulating external information sources that security tools rely on.
They may attempt to influence the system's decision-making process by injecting false information or spreading misinformation across various platforms.
For instance, attackers could artificially enhance a scam website's reputation through fake reviews or manipulated search engine results.
However, ScamFerret's multi-perspective analysis approach, which evaluates websites through web content, DNS records, and search engine results, makes such evasion attempts impractical.
The significant effort and resources required to consistently manipulate multiple information sources across different domains effectively prevent attackers from compromising the system's detection capabilities.

\noindent\textbf{Image-based Scam Attacks.}
Attackers increasingly employ image-based techniques to deceive humans while evading traditional detection systems.
By embedding fraudulent content within images rather than text, attackers can bypass conventional security measures.
While these image-based scams effectively deceive human users, automated detection systems struggle to identify malicious intent in images rather than machine-readable text.
Recent advances in multimodal LLMs like GPT-4V show promise in analyzing visual content for fraud detection, but leveraging these capabilities for comprehensive image-based scam detection remains a future research challenge.

\section{Related Work}
\noindent\textbf{Scam Website Analysis.}
Recent studies have focused on various types of scam websites~\cite{DBLP:conf/ndss/MiramirkhaniSN17,DBLP:conf/ndss/LiYN23,DBLP:conf/www/SrinivasanKMANA18}.
Bitaab et al.'s ``Beyond Phish'' system achieved a 98.34\% detection rate and 1.34\% false positive rate for English scam e-commerce websites~\cite{DBLP:conf/sp/BitaabCOLWAWBSD23}.
Kotzias et al.'s system for detecting fake online shopping websites achieved an F1 score of 0.973~\cite{DBLP:conf/acsac/KotziasRPSB23}.
\textit{Our study extends beyond these by addressing multilingual and multi-type scams.}

\noindent\textbf{Security Task-specific LLMs.}
LLMs for security tasks have gained attention~\cite{alfasi2024unveiling,DBLP:journals/access/KoideNC24}.
Li et al.'s ``KnowPhish Detector'' uses LLMs to extract brand information for phishing detection, achieving a 98.34\% detection rate~\cite{DBLP:conf/uss/LiHDLCOLH24}.
Roy et al. demonstrated LLMs' potential to generate phishing content and proposed a BERT-based detection tool with 96\% accuracy for phishing websites~\cite{DBLP:journals/corr/abs-2310-19181}.
\textit{Our approach differs by leveraging LLMs' text comprehension capabilities for scam website detection.}

\noindent\textbf{LLM-as-a-Judge.}
Recent research has explored LLMs for evaluating LLM-generated content~\cite{DBLP:conf/nips/ZhengC00WZL0LXZ23,DBLP:conf/emnlp/SottanaLZY23}.
Chiang et al. used LLMs for text quality assessment, matching expert human evaluation~\cite{DBLP:conf/acl/ChiangL23}.
Chan et al.'s ``ChatEval'' framework uses multiple LLMs for text generation quality assessment~\cite{DBLP:journals/corr/abs-2308-07201}.
\textit{Our study differs in that it analyzes detection rationale for classifying scam websites, rather than evaluating quality of LLM-generated text.}

\section{Conclusion}
\label{sec:conclusion}
This paper presents ScamFerret, an innovative agent system utilizing LLMs for scam detection without requiring additional training on scam website data.
ScamFerret leverages LLMs' natural language interpretation to identify and analyze nuanced, context-dependent cues indicative of scam websites.
Our evaluation demonstrates high classification accuracy: 0.972 for multiple scam types and 0.993 for multiple languages, providing clear decision rationales.
Unlike traditional machine learning approaches, ScamFerret eliminates the need for additional training data, complex feature engineering, and frequent model updates.
It autonomously collects information based on scam characteristics provided in natural language, enabling effective detection without conventional constraints.
This work advances LLM applications in cybersecurity and opens new research directions.

\bibliographystyle{splncs04}
\bibliography{mybib}

\begin{thebibliography}{10}
\providecommand{\url}[1]{\texttt{#1}}
\providecommand{\urlprefix}{URL }
\providecommand{\doi}[1]{https://doi.org/#1}

\bibitem{fraud_distribtuion}
{ David Janssen}: How do the world’s biggest countries deal with online fraud? \url{https://vpnoverview.com/internet-safety/cybercrime/how-countries-deal-with-online-fraud-and-cybercrime/} (2024)

\bibitem{alfasi2024unveiling}
Alfasi, D., Shapira, T., Barr, A.B.: Unveiling hidden links between unseen security entities (2024)

\bibitem{DBLP:conf/sp/BitaabCOLWAWBSD23}
Bitaab, M., et~al.: Beyond phish: Toward detecting fraudulent e-commerce websites at scale. In: Proc. IEEE SP (2023)

\bibitem{scam_reason_2}
{CBS News}: Cryptocurrency fraud is now the riskiest scam for consumers, according to bbb. \url{https://www.cbsnews.com/news/crypto-scam-risk-bbb-report/} (2024)

\bibitem{DBLP:journals/corr/abs-2308-07201}
Chan, C., et~al.: Chateval: Towards better llm-based evaluators through multi-agent debate  (2023)

\bibitem{DBLP:conf/acl/ChiangL23}
Chiang, D.C., et~al.: Can large language models be an alternative to human evaluations? In: Proc. ACL (2023)

\bibitem{curlie}
{Curlie.org}: Curlie - the collector of urls. \url{https://curlie.org/} (2024)

\bibitem{DBLP:conf/IEEEares/DooremaalBAZ21}
van Dooremaal, B., et~al.: Combining text and visual features to improve the identification of cloned webpages for early phishing detection. In: Proc. {ARES} (2021)

\bibitem{scam_reason_1}
{Fashion United}: Online shopping fraud in the uk is a 2.3bn pounds crisis. \url{https://fashionunited.uk/news/retail/online-shopping-fraud-in-the-uk-is-a-2-3bn-pounds-crisis/2024111278529} (2024)

\bibitem{fbi_report}
{FBI}: Fbi releases internet crime report. \url{https://www.ic3.gov/Media/PDF/AnnualReport/2023_IC3Report.pdf} (2024)

\bibitem{DBLP:journals/corr/abs-2312-10997}
Gao, Y., et~al.: Retrieval-augmented generation for large language models: {A} survey  (2023)

\bibitem{gemini}
{Google DeepMind}: Gemini. \url{https://deepmind.google/technologies/gemini/} (2024)

\bibitem{DBLP:conf/imc/HaoFP11}
Hao, S., et~al.: Monitoring the initial {DNS} behavior of malicious domains. In: Proc. ACM {IMC} (2011)

\bibitem{DBLP:conf/asiaccs/KimCKDSAD21}
Kim, D., et~al.: Security analysis on practices of certificate authorities in the {HTTPS} phishing ecosystem. In: Proc. ACM ASIACCS (2021)

\bibitem{DBLP:journals/access/KoideNC24}
Koide, T., et~al.: Chatphishdetector: Detecting phishing sites using large language models. {IEEE} Access  (2024)

\bibitem{DBLP:conf/acsac/KotziasRPSB23}
Kotzias, P., et~al.: Scamdog millionaire: Detecting e-commerce scams in the wild. In: Proc. ACSAC (2023)

\bibitem{langchain}
{LangChain}: Langchain. \url{https://www.langchain.com/} (2024)

\bibitem{DBLP:conf/ndss/LiYN23}
Li, X., et~al.: Double and nothing: Understanding and detecting cryptocurrency giveaway scams. In: Proc. NDSS (2023)

\bibitem{DBLP:conf/uss/LiHDLCOLH24}
Li, Y., et~al.: Knowphish: Large language models meet multimodal knowledge graphs for enhancing reference-based phishing detection. In: Proc. {USENIX} Security (2024)

\bibitem{DBLP:conf/uss/LinLDNCLSZD21}
Lin, Y., et~al.: Phishpedia: {A} hybrid deep learning based approach to visually identify phishing webpages. In: Proc. {USENIX} Security (2021)

\bibitem{DBLP:conf/eurosp/LiuPVP23}
Liu, J., et~al.: Understanding, measuring, and detecting modern technical support scams. In: Proc. IEEE EuroSP (2023)

\bibitem{eng_beyond}
{mbitaab}: beyondphish. \url{https://github.com/mbitaab/beyondphish} (2024)

\bibitem{azure_openai}
{Microsoft Azure}: Azure openai service – advanced language models. \url{https://azure.microsoft.com/en-us/products/ai-services/openai-service/} (2024)

\bibitem{DBLP:conf/ndss/MiramirkhaniSN17}
Miramirkhani, N., et~al.: Dial one for scam: {A} large-scale analysis of technical support scams. In: Proc. NDSS (2017)

\bibitem{jpn_onlineshopping_fraud}
{Neoblood Corporation}: Disclosure of information on fake website. \url{https://www.neo-blood.co.jp/} (2024)

\bibitem{eng_tss_fraud1}
{NISLabUGA}: Tss esp23. \url{https://github.com/NISLabUGA/TSS_ESP23} (2024)

\bibitem{eng_tss_fraud2}
{NOLA Defense}: Nola defense. \url{https://www.noladefense.net/} (2024)

\bibitem{DBLP:conf/uss/OestZWNBZTDA20}
Oest, A., et~al.: Sunrise to sunset: Analyzing the end-to-end life cycle and effectiveness of phishing attacks at scale. In: Proc. {USENIX} Security (2020)

\bibitem{playwright}
{Playwright}: Fast and reliable end-to-end testing for modern web app. \url{https://playwright.dev/} (2024)

\bibitem{DBLP:conf/ndss/PochatGTKJ19}
Pochat, V.L., et~al.: Tranco: {A} research-oriented top sites ranking hardened against manipulation. In: Proc. NDSS (2019)

\bibitem{reddit_api}
{reddit inc.}: reddit.com: api documentation. \url{https://www.reddit.com/dev/api/} (2024)

\bibitem{DBLP:journals/corr/abs-2310-19181}
Roy, S.S., et~al.: From chatbots to phishbots? - preventing phishing scams created using chatgpt, google bard and claude  (2023)

\bibitem{DBLP:journals/corr/abs-2401-09824}
Saad, B.A.M., et~al.: Conning the crypto conman: End-to-end analysis of cryptocurrency-based technical support scams  (2024)

\bibitem{eng_onlineshopping_fraud1}
{ScamGuard™}: Listings. \url{https://scamguard.com/reviews/} (2024)

\bibitem{crtsh}
{Sectigo}: crt.sh | certificate search. \url{https://crt.sh/} (2024)

\bibitem{DBLP:conf/emnlp/SottanaLZY23}
Sottana, A., et~al.: Evaluation metrics in the era of {GPT-4:} reliably evaluating large language models on sequence to sequence tasks. In: Proc. EMNLP (2023)

\bibitem{DBLP:conf/www/SrinivasanKMANA18}
Srinivasan, B., et~al.: Exposing search and advertisement abuse tactics and infrastructure of technical support scammers. In: Proc. {WWW} (2018)

\bibitem{DBLP:conf/ccs/TangML0022}
Tang, S., et~al.: Clues in tweets: Twitter-guided discovery and analysis of {SMS} spam. In: Proc. ACM CCS (2022)

\bibitem{tavily}
{Tavily}: Tavily. \url{https://tavily.com/} (2024)

\bibitem{eng_onlineshopping_fraud2}
{The Scam Directory}: The scam directory. \url{https://scam.directory/} (2024)

\bibitem{trustpilot}
{Trustpilot}: Trustpilot reviews: Experience the power of customer reviews. \url{https://www.trustpilot.com/} (2024)

\bibitem{DBLP:conf/imc/VallinaPFPGBHTV20}
Vallina, P., et~al.: Mis-shapes, mistakes, misfits: An analysis of domain classification services. In: Proc. ACM {IMC} (2020)

\bibitem{virustotal}
VirusTotal: Virustotal (2024), \url{https://www.virustotal.com/}

\bibitem{ger_onlineshopping_fraud}
{Watchlist Internet}: Fraudulent online stores. \url{https://www.watchlist-internet.at/liste-betruegerischer-shops/} (2024)

\bibitem{DBLP:journals/corr/abs-2302-11382}
White, J., et~al.: A prompt pattern catalog to enhance prompt engineering with chatgpt  (2023)

\bibitem{twitter_api}
{X Corp}: Twitter api | products | twitter developer platform. \url{https://developer.twitter.com/en/products/twitter-api} (2024)

\bibitem{eng_cryptoscam}
{Xigao Li}: Double and nothing: Understanding and detecting cryptocurrency giveaway scams. \url{https://double-and-nothing.github.io/} (2024)

\bibitem{DBLP:conf/iclr/YaoZYDSN023}
Yao, S., et~al.: React: Synergizing reasoning and acting in language models. In: Proc. ICLR (2023)

\bibitem{DBLP:conf/nips/ZhengC00WZL0LXZ23}
Zheng, L., et~al.: Judging llm-as-a-judge with mt-bench and chatbot arena. In: Proc. NeurIPS (2023)

\bibitem{DBLP:conf/iclr/ZhouMHPPCB23}
Zhou, Y., et~al.: Large language models are human-level prompt engineers. In: Proc. ICLR (2023)

\end{thebibliography}

\appendix

\end{document}